# Segmentation and Risk Score Prediction of Head and Neck Cancers in PET/CT Volumes with 3D U-Net and Cox Proportional Hazard Neural Networks

Fereshteh Yousefirizi[1][0000-0001-5261-6163], Ian Janzen[1], Natalia Dubljevic[2], Yueh-En Liu[3], Chloe Hill[4], Calum MacAulay[1,2,5], Arman Rahmim[1,2,6][0000-0002-9980-2403]

[1] Department of Integrative Oncology, BC Cancer Research Institute, Vancouver, Canada
[2] Department of Physics and Astronomy, University of British Columbia, Vancouver, Canada
[3] Combined Major in Science, University of British Columbia, Vancouver, BC, Canada
[4] Faculty of Applied Sciences, Simon Fraser University, Burnaby, Canada
[5] Department of Pathology, University of British Columbia, Vancouver, Canada
[6] Department of Radiology, University of British Columbia, Vancouver, Canada
frizi@bccrc.ca

**Abstract.** We utilized a 3D nnU-Net model with residual layers supplemented by squeeze and excitation (SE) normalization for tumor segmentation from PET/CT images provided by the Head and Neck Tumor segmentation challenge (HECKTOR). Our proposed loss function incorporates the Unified Focal and Mumford-Shah losses to take the advantage of distribution, region, and boundary-based loss functions. The results of leave-one-out-center-cross-validation performed on different centers showed a segmentation performance of 0.82 average Dice score (DSC) and 3.16 median Hausdorff Distance (HD), and our results on the test set achieved 0.77 DSC and 3.01 HD. Following lesion segmentation, we proposed training a case-control proportional hazard Cox model with an MLP neural net backbone to predict the hazard risk score for each discrete lesion. This hazard risk prediction model (CoxCC) was to be trained on a number of PET/CT radiomic features extracted from the segmented lesions, patient and lesion demographics, and encoder features provided from the penultimate layer of a multi-input 2D PET/CT convolutional neural network tasked with predicting time-to-event for each lesion. A 10-fold cross-validated CoxCC model resulted in a c-index validation score of 0.89, and a c-index score of 0.61 on the HECKTOR challenge test dataset.

**Keywords:** Head and neck cancer, PET-CT, Segmentation, Unified focal loss, Cox regression

## 1. Introduction

Head and neck (H&N) cancer is the fifth most common cancer diagnosed worldwide and the eighth most common cause of cancer death [1]. While standard therapies combining radiation and chemotherapy are highly effective, they rely heavily on manually

---

The first two authors contributed equally.

generated contours of tumor volumes from medical images. Segmentation is also a crucial bottleneck towards radiomics analysis and prognostication pipelines. This is a labor intensive and time-consuming task that unavoidably suffers from intra- and inter-observer biases [2, 3]. AI approaches to tumor segmentation continue to grow in popularity, and have demonstrated potential for identification of head and neck tumors in PET and CT image tasks. Previous network approaches range from a simple U-net model, to 3D V-nets and 3D deep networks.

Radiomics are hand-engineered statistical features that are predominantly calculated from masked segmentations acquired from medical imaging domains. Moreover, they refer to the use of image analysis to quantify image descriptors, called radiomic features, that are often considered imperceptible to the human eye. Radiomic features typically calculated from PET and CT images are often described as shape, textural, and intensity features. Studies have shown that radiomics help clinicians bridge the gap between quantitative and qualitative image analysis to help them understand the biological processes underlying the image phenomena [4].

Clinically relevant features extracted from PET and CT images are powerful tools for classifying pathological behaviors. Many studies have shown the importance of relevant radiomics features in assessing images and assigning appropriate treatment pathways [2,4,18]. Identifying relevant radiomic features requires a robust and algorithmic feature selection process. These methods are required to remove potential evaluation bias while identifying discriminating features that are useful for clinicians to help them quantize lesion characteristics.

The data provided for this study is well suited for a survival analysis and time-to-event prediction task. We explored the use of a number of discrete and continuous time Cox regression models to predict either hazard risk scores or time-to-event predictions for individual head and neck cancers segmented lesions. These models relied on a Multilayer Perception (MLP) neural network backbone, as an extension of the Cox regression model [5], to predict this task. Further explanation of the features that these models were trained on will be elucidated on in the proposed methods.

In the current study, we propose a 3D network for bi-modal PET/CT segmentation based on a 3D nnU-net model with squeeze and excitation (SE) modules and a hybrid loss function (distribution, region and boundary based). Using this network, we constructed an automated radiomic extraction pipeline to generate potentially useful features for a risk prediction score provided by a Cox proportional hazard model.

In the following sections, we first introduce the data provided by the MICCAI 2020 HEad and neCK TumOR segmentation and outcome prediction (HECKTOR) challenge [6, 7]. Our proposed methods and training scheme are additionally explained. The results are then presented, followed by discussion and conclusion.

4## 2. Material and Methods

### 2.1 Dataset

**2.1.1 Description.** Operating under the confines of the MICCAI 2021 HECKTOR Data Challenge, we were provided with 224 discrete PET/CT and lesion mask volumes that contained head and neck cancers collected from 5 different sites to train and validate deep learning models. An external cohort of PET/CT and lesion mask volumes (N=101), collected from two different sites were also provided for these models. This external cohort assessed the performance of each task and gauged its generalizability. For the segmentation task, training and validation splits were executed as a leave-one-out-center-cross-validation to assess model performance on the generalizability on out-of-sample data and random splitting was done. For the risk score prediction task, training and validation splits were done with pseudo-random training and validation splits (90:10 - training: validation) with 10-fold cross-validated. Prior to pseudo-random splitting, we manually selected an "external" validation set of 21 volumes, with an approximate distribution of disease progression and progression free survival of days (15:6 - non event:event). This was done to better assess model ranking ability on out-of-sample data.

**2.1.2 Preprocessing.** Based on the directives and bounding boxes provided by HECKTOR challenge organizers, the PET and CT images were resampled (by trilinear interpolation), and then cropped, to the resolution of $1\times1\times1$ mm$^3$. We clipped the intensities of CT images in the range of $[-1024, 1024]$ Hounsfield Units and then resampled to within a $[-1, 1]$ range. PET images were normalized by Z-score normalization. For the segmentation task, the volumes were resampled back to their original resolutions before verifying their statistical performance. In addition to mirroring (on the axial plane) and rotation (in random directions) for data augmentation, we utilized scaling (with a random factor between 0.8 and 1.2) and elastic deformations to increase the diversity in tumor size and shape.

### 2.2 Proposed methods

**2.2.1 Segmentation.** As the backbone network for medical image segmentation, 3D U-Net [8] and 3D nnU-net [9] have gained much attention among convolutional neural networks (CNNs) due to their good performances. However, the upsampling process involves the recovery of spatial information, which is difficult without taking the global information into consideration [10]. The squeeze & excitation (SE) modules are defined to 'squeeze' along the spatial domain and 'excite' along the channels. SE modules help the model to highlight the meaningful features and suppress the weak ones. CNNs with SE modules frequently achieve top performance, across various challenges (ILSVRC 2017 image classification [10] and Head and Neck Tumor segmentation challenge (HECKTOR 2020) [11]). We utilized the 3D nnU-Net with SE modules after encoder and decoder blocks as the recommended architecture by Roy et al [12].

Using different loss functions has been shown to affect the performance, robustness and convergence of the segmentation network [13]. Distribution based losses (e.g. cross entropy, Focal loss [14]), region based losses (e.g. Dice), boundary based loss (e.g. Mumford-Shah [15]) or any of their combinations make up the main approaches for loss functions in medical image segmentation tasks. Hybrid loss functions have shown better performance [13, 16] e.g. the sum of cross entropy and Dice similarity coefficient (DSC) proposed by Taghanaki et al. [17] or the Unified Focal loss introduced by Yeung et al. [13]that combined Focal and Dice (the minus of DSC) loss. We used a 3D nnU-Net with SE modules (Fig. 1) and utilized a hybrid loss function that is a combination of the distribution based, region based and boundary based loss functions. The model was trained for 400 epochs using Adam optimizer on two NVIDIA Tesla V100 GPUs 16 GB with a batch size of 2. We used the cosine-annealing schedule to reduce the learning rate from $10^{-3}$ to $10^{-6}$ within every 25 epochs.

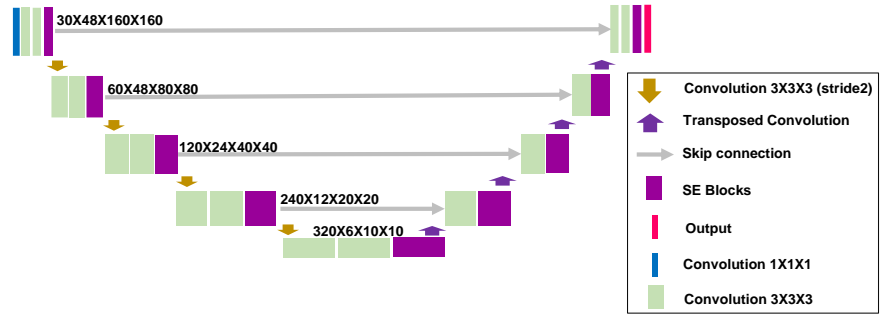

**Fig. 1.** 3D nnU-Net model for segmentation with SE modules

**2.2.2 Radiomic Extraction and Selection.** Radiomic features have been shown to improve patient outcomes and assist physicians by quantifying phenotypical behaviors in images as basic statistical inference. We propose utilizing the Pyradiomics Python package[18] to extract radiomic features from both the PET and CT volumes, using provided lesion masks (either ground truth, or from the segmentation model). We extracted approximately 3000 radiomic features from both modalities using this methodology.

To isolate the most discriminating features, we utilized the scikit-learn python package's feature selection algorithms[19]. We performed a robust search grid through the available classification-criterion and feature selection algorithms as an exhaustive method to identify the most discriminating features. Through scikit-learn, we identified four feature algorithms: Sequential Forward Feature Selection (SFS; forwards and backwards), Extra Trees Classifier (ETC), Recursive Feature Elimination (RFE). In addition, we exposed our results to the use of the Maximum Relevance - Minimum Redundancy (MRMR) feature selection algorithm [20], to verify discrete feature discriminating viability. For each feature selection method, we use one of the following criterion models: Linear Discriminant Analysis (LDA), K-Nearest Neighbors (KNN), Bagging methods, Gradient Boosting (GradBoost), eXtreme Gradient Boosting (XGBoost),

Support Vector Machines (SVM), Classification and Regression Trees (CART), Quadratic Discriminant Analysis (QDA), and Linear Regression (LR). Each criterion model was provided with the approx. 3000 radiomic features and the ground truth as the binarized indicated progression status of each patient. From there, we searched and logged the top 1, 5, 10, 25, 50, and 100 discriminating features using each combination of feature selection methods and criterion models.

**2.2.3 Encoder Feature Extraction.** Inspired by the segmentation aspect of the HECKTOR 2021 challenge, we decided that encoded features ought to be implemented in the time-to-event prediction task. We used a multi-input model 2D convolution neural network (CNN) that would be trained to predict the progression free survival, measured in days (PFS days). The CNN would predict this time-to-event given the PET and CT slices that correspond to an axial slice that includes a ground-truth mask. However, provided this is a censored data task, we implemented a custom loss function to train the model that respects an over-estimation of PFS days for non-progression events. We have colloquially referred to this function as a One Way Penalized Survival loss (Eq 1). Here Pr refers to the progression of the patient (binary value), and Mean Squared Error (MSE) is used as a placeholder example error function to update the model weights during training. This custom loss function ensured that a model would not be penalized for over predicting the number of PFS days for censored data, thus gaining no weight updates on censored data beyond the last observation. To further restrict how the model may learn, batch training was not implemented. This simplistic CNN was trained for 5 epochs, with an Adam optimizer and a learning rate of 0.001. This model achieved a MSE of 483 days between the PFS days and predicted PFS days for the validation set.

$$loss(PFS_{pred}, Pr, PFS_{GT}) = \begin{cases} 0 & if\ Pr = 0\ and\ PFS_{pred} > PFS_{GT} \\ 1 & else \end{cases} \quad (1)$$

As we were provided PET/CT volumes, we extracted a dataset using the same preprocessing methods outlined in section "description". The penultimate layer of 5X fold validation multi-input 2D CNN model trained under this regime was isolated and the features of that layer, averaged per the number of slices where an axial mask was present in the respective volume, were then to be accessed for training our Cox proportional hazard model.

**2.2.4 Outcome Prediction.** The intention of this experiment is to build on the nested case-control studies conducted by Langholz et al. [21] We extend their work by utilizing the PyTorch framework [22] to construct a case-control Cox proportional hazard model that is able to predict a proportional hazard risk score. We will perform this task by combining the three sources of features listed in section Radiomic Extraction and Selection and Outcome Prediction alongside one-hot-encoded patient demographic features.

These three sources of features are used as the inputs for a case-control Cox Regression model (CoxCC) with a MLP neural network backbone [5] using the pycox python package. This case-control proportional hazard Cox regression model predicts the hazard risk score per patient using a linear combination of patient demographics, selected radiomic features, and encoder features generated from the 2D CNN model. A visualization of the entire feature extraction-to-CoxCC model training block diagram can be seen in Fig.2. The overall CoxCC model is trained using the loss function described by Kvamme et al. [5] for case-control models, an Adam optimizer, learning rate of 0.0024, a batch size of 32, and an early stopping method that monitored the validation loss (tolerance of 3 epochs) to combat overtraining by saving the optimized model weights.

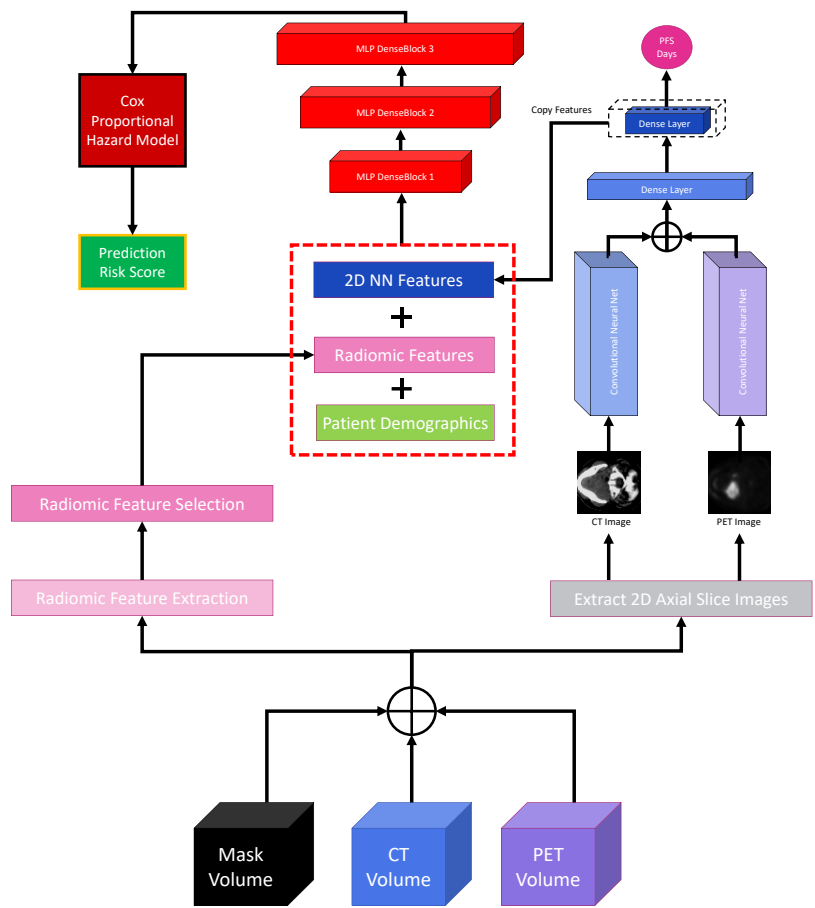

**Fig. 2.** Overall MICCAI 2021 HECKTOR Challenge Task 2 data extraction and model training pipeline

## 3. Results

**3.1 Segmentation Results.** As Table 1 presents our 3D nnU-Net model with SE modules outperformed 3D nnU-Net using the same loss function for training. The best segmentation performance observed using the proposed compound loss function incorporates the distribution, region and boundary based loss functions (Unified Focal loss and Mumford-Shah).

**Table 1.** Segmentation performance in different models/losses

| Method | Loss | DSC (mean) | HD (median) |
| --- | --- | --- | --- |
| 3D nnU-Net | Unifocal | 0.68±0.19 | 11.07 |
| 3D nnU-Net | Hybrid | 0.72± 0.23 | 9.32 |
| 3D nnU-Net (SE) | Unifocal | 0.79± 0.16 | 6.4 |
| 3D nnU-Net (SE) | Hybrid | 0.82±0.12 | 4.09 |

In the validation phase, the best DSC value was achieved on the data from CHGJ center (n=55) and the best performance in terms of HD metric was received on data from CHUS center (n=72) (Table 2). The model demonstrated the average DSC of 0.82± 0.12 on leave-one-out-center-cross-validation and 0.84± 0.28 on random split that showed no significant difference.

**Table 2.** Segmentation performance on different leave-one-out-center-cross-validation

| Centers | DSC (mean ± std) | HD (median) |
| --- | --- | --- |
| CGHJ (n=55) | 0.8655 ± 0.0627 | 3 |
| CHMR (n=18) | 0.7919 ± 0.1396 | 5.5495 |
| CHUM(n=56) | 0.7951 ± 0.1327 | 3.1623 |
| CHUP(n=23) | 0.7888 ± 0.1214 | 5 |
| CHUS (n= 72) | 0.8101 ± 15.79 | 2.4495 |

Our results on the test set were calculated using an ensemble of ten models i.e. five trained and validated on center-based splitting and five trained on random splitting. For leave-one-center-out cross-validation, we used images from four centers for training and data from the fifth center was used for validation. The predicted masks on test data were calculated by averaging the predictions by the above-mentioned ten models (threshold value=0.5).

**3.2 Survival analysis.** We used the concordance index (c-index) to assess survival model performance on this censored dataset. We performed 10-fold cross-validated on these models to verify model efficacy. Thus, when assessing c-index scores on our "external" validation set and the test set, we took the median prediction from these 10 models for each segmented lesion to generate a single prediction risk score from our CoxCC model method.

A grid search involving all criterion models and feature selection methods identified 192 unique combinations of radiomic feature to train a CoxCC model with. We also tested the prediction capability of the CoxCC model against a regular Cox Proportional Hazard (CoxPH) model using the same radiomic features. We identified the boost to predicted risk score capability with the addition of the features described in section "Radiomic Extraction and Selection" to help elucidate final model selection choice. The results of model selection tabulated in Table 3 and an example plot comparing each model against other feature selection methodologies in Fig. 3 and Fig 4.

**Table 3.** Identifying the top performing models for Task 2. Acronyms for Feature Selection Method can be found in section "Radiomic Extraction and Selection"

| Model Type | CoxPH | CoxCC | CoxPH | CoxCC | CoxPH | CoxCC | CoxPH | CoxCC | CoxPH | CoxCC |
|---|---|---|---|---|---|---|---|---|---|---|
| # of Radiomic features | 5 | 5 | 10 | 10 | 25 | 25 | 50 | 50 | 100 | 100 |
| Feature Selection Method | LDA + ETC | QDA + ETC | LDA + SFS F | SVM + ETC | KNN + ETC | GradBoost + ETC | LR + ETC | GradBoost + SFS F | LR + ETC | GradBoost + SFS F |
| Training set C-index | 0.811 | 0.883 | 0.743 | 0.870 | 0.774 | 0.867 | 0.769 | 0.891 | 0.758 | 0.884 |
| Validation set C-index (10X fold) | 0.744 | 0.871 | 0.638 | 0.891 | 0.646 | 0.847 | 0.649 | 0.886 | 0.612 | 0.883 |
| Ext Validation set C-index (median) | 0.813 | 0.854 | 0.865 | 0.888 | 0.843 | 0.888 | 0.876 | **0.910** | 0.753 | **0.910** |
| Test Set C-index (mean) | - | - | - | - | - | **0.576** | - | **0.604** | - | **0.612** |

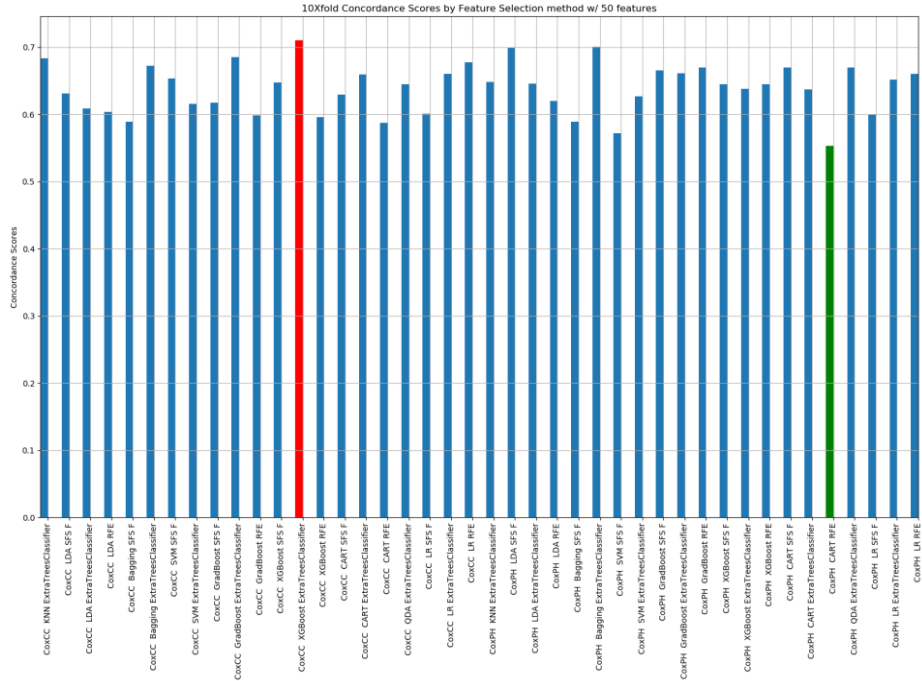

**Fig.3.** Close-case Cox (CoxCC) regression method against the Cox Proportional Hazard model (CoxPH) with the validation set c-index score when the features from the custom 2D Neural Network are not introduced during training. All models trained with 50 radiomic features, selected by the routines documented in the x-axis titles and 5 patient demographic features. Compared to Figure 4, one can see a boost to outcome prediction performance with the addition of the features described in section 2.2.3. [Color: Highest c-index score marked in red, lowest c-index score marked in green].

## 4. Discussion

Our segmentation results show that utilizing SE modules improves the performance of 3D nnU-net model for segmentation. On the other side, a compound loss function composed of distribution, region and boundary based losses also improves the segmentation performance of a 3D nnUnet model with SE modules.

An exhaustive search for the most discriminating radiomic features for predicting hazard risk scores of head and neck cancer patients proved to not have much of a marked effect on these static model architectures. Generally, the features selected were a healthy mix of CT- and PET-based radiomic features as well. Observing Fig. 3, one can conclude that there is little discernible difference in model performance between the CoxCC and CoxPH methods despite additional radiomic features to train with. An average validation set c-index score of 0.62 for the myriad of CoxCC and CoxPH models trained on solely radiomic features and patient demographics does not make for an ideal predictor. This observation is neatly contrasted by the performance of the CoxCC model vs CoxPH model when the 2D NN encoded features are able to be trained with. In figure 4 and Table 3, there is a clear boost, approx. +0.09 c-index, to prediction ability on the validation set with the addition of these encoded features during training. Moreover, we see excellent model performance on the external validation set from the CoxCC models with more than 25 radiomic features, seen in Table 3. Particularly noteworthy are the CoxCC models' ability to rank patient risk scores on data that is in the training sets time-to-event distribution.

However, the ranking ability for these models is severely impacted by predicting risk scores for lesions that are out-of-sample data distribution. In this case, prediction on the test set resulted in a drop of approximately 0.30 c-index points from the external validation set to the test set. It is a fair assumption to suppose that these models are overtrained on the 2D NN features from the 5 sites that the training data was collected from. These models showcase a poor ability to generalize currently. Thus, they would require more rigorous regularization techniques, a more exhaustive neural network architecture search, and the introduction of more data samples with which to train.

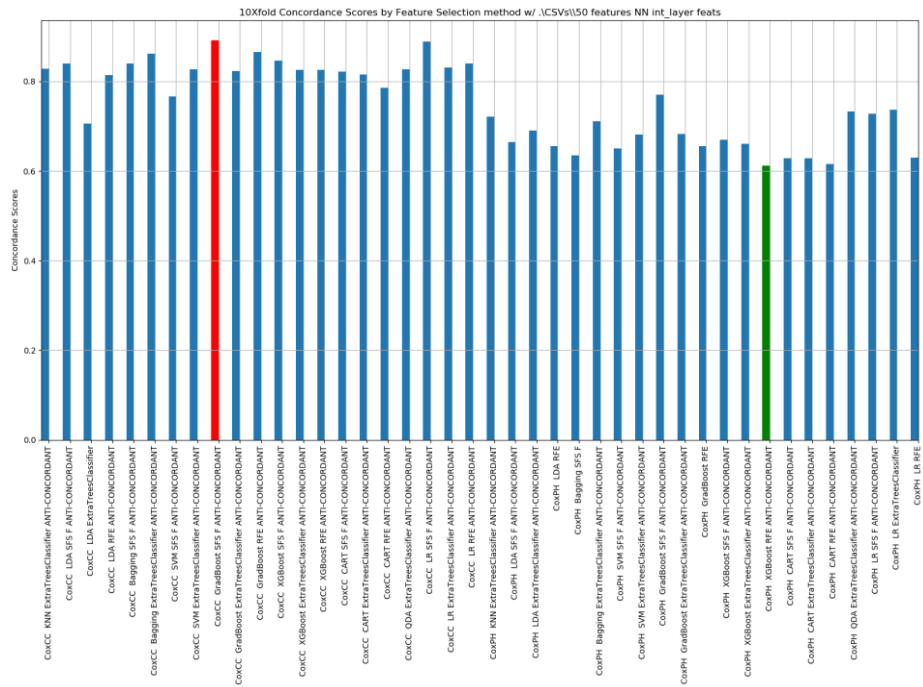

**Fig. 4.** Showcasing the boost to c-index performance with the close-case Cox (CoxCC) regression method against the Cox Proportional Hazard model (CoxPH) and the validation set c-index score. All models trained with 50 radiomic features, selected by the routines documented in the x-axis tick titles, 5 patient demographic features, and 128 2D Neural Network activation layer features extracted from the penultimate layer. [Color: highest c-index score marked in red, lowest c-index score marked in green.]

## 5. Conclusion

The segmentation and survival prediction for head and neck cancers are both difficult tasks. The MICCAI 2021 HECKTOR Challenge provided the groundwork to test our methodologies on the execution of these tasks. SE modules help to improve the performance of nnU-Net models for PET and PET/CT image segmentation. Hybrid loss functions that take into consideration the distribution, region and boundary improve the segmentation performance. The other facet of the HECKTOR challenge, providing an accurate patient risk score, proved to be just as troublesome. Radiomic features were extracted from the PET and CT volumes, and then algorithmically selected. The radiomics were combined with patient demographics and encoded features of a 2D NN trained to predict patient survival in days. This combination of features were trained on by a case-control Cox regression model to estimate an overall patient risk score. While this model achieved impressive results on in-distribution data, it failed to generalize well on out-of-distribution data from other PET/CT collection sites. In summation, head and neck cancers are difficult to segment and can confound patient risk models. Yet they prove to be an intriguing and insightful challenge for researchers, clinical or otherwise, to diagnose and help keep patients alive for longer.

## Acknowledgement:

This project was in part supported by the Natural Sciences and Engineering Research Council of Canada (NSERC) Discovery Grant RGPIN-2019-06467, and the Canadian Institutes of Health Research (CIHR) Project Grant PJT-173231.